\documentclass[twocolumn,floatfix,prb]{revtex4}
\usepackage{graphicx}
\begin{document}
\title{\Huge\bf\sc Quantum Shot Noise}
\author{\large\bf Fluctuations in the flow of electrons signal the transition\\
from particle to wave behavior.\\
{\tt Published in revised form in Physics Today, May 2003, page 37.}\bigskip}
\affiliation{\large Carlo Beenakker \& Christian Sch\"{o}nenberger}
\thanks{
{\sc Carlo Beenakker} {\em is at the Instituut-Lorentz of
Leiden University (The Netherlands).} {\sc Christian Sch\"{o}nenberger} {\em is
at the Physics Department of the University of Basel (Switzerland).}
}
\maketitle

\noindent
``The noise is the signal'' was a saying of Rolf Landauer, one of the
founding fathers of mesoscopic physics. What he meant is that fluctuations
in time of a measurement can be a source of information that is not
present in the time-averaged value. A physicist may actually delight in noise.

Noise plays a uniquely informative role in connection with the
particle-wave duality. It was Albert Einstein who first realized (in
1909) that electromagnetic fluctuations are different if the energy is
carried by waves or by particles. The magnitude of energy fluctuations
scales linearly with the mean energy for classical waves, but it scales
with the square root of the mean energy for classical particles. Since
a photon is neither a classical wave nor a classical particle, the
linear and square-root contributions coexist. Typically, the square-root
(particle) contribution dominates at optical frequencies, while the linear
(wave) contribution takes over at radio frequencies. If Newton could
have measured noise, he would have been able to settle his dispute with
Huygens on the corpuscular nature of light --- without actually needing
to observe an individual photon. Such is the power of noise.

The diagnostic property of photon noise was further developed in the
1960's, when it was discovered that fluctuations can tell the difference
between the radiation from a laser and from a black body: For a laser
the wave contribution to the fluctuations is entirely absent, while it
is merely small for a black body. Noise measurements are now a routine
technique in quantum optics and the quantum mechanical theory of photon
statistics (due to Roy Glauber) is textbook material.

Since electrons share the particle-wave duality with photons, one might
expect fluctuations in the electrical current to play a similar diagnostic
role. Current fluctuations due to the discreteness of the electrical
charge are known as ``shot noise''. Although the first observations of
shot noise date from work in the 1920's on vacuum tubes, our quantum
mechanical understanding of electronic shot noise has progressed more
slowly than for photons. Much of the physical information it contains
has been appreciated only recently, from experiments on nanoscale
conductors.\cite{Bla00}

\section*{\large\bf Types of electrical noise}
\noindent
Not all types of electrical noise are informative. The fluctuating voltage
over a conductor in thermal equilibrium is just noise. It tells us
only the value of the temperature $T$. To get more out of noise one
has to bring the electrons out of thermal equilibrium. Before getting
into that, let us say a bit more about thermal noise --- also known as
``Johnson-Nyquist noise'' after the two physicists who first studied it
in a quantitative way.

Thermal noise extends over all frequencies up to the quantum limit at
$kT/h$. In a typical experiment one filters the fluctuations in a band
width $\Delta f$ around some frequency $f$. Thermal noise then has
an electrical power of $4kT\Delta f$, independent of $f$ (``white''
noise). One can measure this noise power directly by  the amount of
heat that it dissipates in a cold reservoir. Alternatively, and this is
how it is usually done, one measures the (spectrally filtered) voltage
fluctuations themselves. Their mean squared is the product $4kTR\Delta f$
of the dissipated power and the resistance $R$.

Theoretically, it is easiest to describe electrical noise in terms of
frequency-dependent current fluctuations $\delta I(f)$ in a conductor
with a fixed, nonfluctuating voltage $V$ between the contacts. The
equilibrium thermal noise then corresponds to $V=0$, or a short-circuited
conductor. The spectral density $S$ of the noise is the mean-squared
current fluctuation per unit band width:
\begin{equation}
S(f)=\langle \delta I(f)^{2}\rangle/\Delta f.\label{Sdef}
\end{equation}
In equilibrium $S=4kTG$, independent of frequency. If a voltage $V\neq
0$ is applied over the conductor, the noise rises above that equilibrium
value and becomes frequency dependent.

At low frequencies (typically below 10~kHz) the noise is dominated by
time-dependent fluctuations in the conductance, arising from random
motion of impurities. It is called ``flicker noise'', or ``$1/f$
noise''  because of the characteristic frequency dependence. Its spectral
density varies quadratically with the mean current $\bar{I}$. At higher
frequencies the spectral density becomes frequency independent and
linearly proportional to the current. These are the characteristics of
shot noise.

The term ``shot noise'' draws an analogy between electrons and the small
pellets of lead that hunters use for a single charge of a gun. The analogy
is due to Walter Schottky, who predicted in 1918 that a vacuum tube
would have two intrinsic sources of time-dependent current fluctuations:
Noise from the thermal agitation of electrons (thermal noise) and noise
from the discreteness of the electrical charge (shot noise).

In a vacuum tube, electrons are emitted by the cathode randomly and
independently. Such a Poisson process has the property that the mean
squared fluctuation of the number of emission events is equal to the
average count. The corresponding spectral density equals $S=2e\bar{I}$.
The factor of 2 appears because positive and negative frequencies
contribute identically.

\section*{\large\bf Measuring the unit of transferred charge}
\noindent

Schottky proposed to measure the value of the elementary charge
from the shot noise power, perhaps more accurately than in the oil
drop measurements which Robert Millikan had published a few years
earlier. Later experiments showed that the accuracy is not better than a
few percent, mainly because the repulsion of electrons in the space around
the cathode invalidates the assumption of independent emission events.

It may happen that the granularity of the current is not the elementary
charge. The mean current can not tell the difference, but the noise can:
$S=2q\bar{I}$ if charge is transferred in independent units of $q$. The
ratio $F=S/2e\bar{I}$, which measures the unit of transferred charge,
is called the ``Fano factor'', after Ugo Fano's 1947 theory of the
statistics of ionization.

\begin{figure}
\includegraphics[width=.8\linewidth]{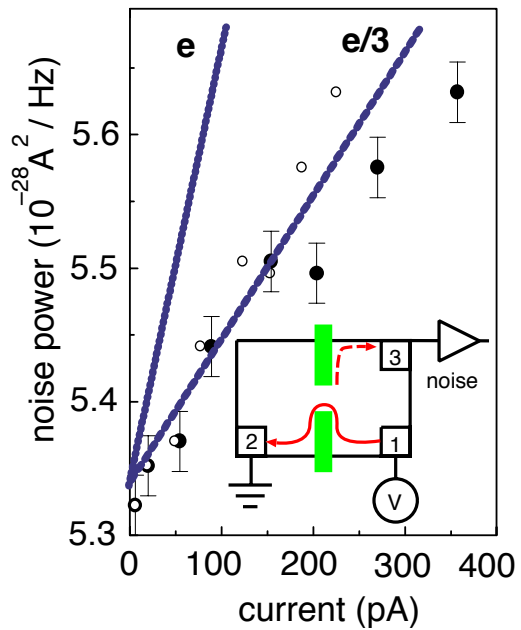}
\caption{
Current noise for tunneling across a Hall bar on the 1/3 plateau
of the fractional quantum Hall effect. The slopes for $e/3$ charge
quasiparticles and charge $e$ electrons are indicated. The data points
with error bars (from the experiment of Saminadayar et al.\cite{Sam97})
are the measured values at 25~mK, the open circles include a correction
for finite tunnel probability. The inset shows schematically the setup
of the experiment. Most of the current flows along the lower edge of
the Hall bar from contact 1 to contact 2 (solid red line), but some
quasiparticles tunnel to the upper edge and end up at contact 3
(dashed). The tunneling occurs predominantly at a narrow constriction,
created in the Hall bar by means of a split gate electrode (shown in
green). The current at contact 3 is first spectrally filtered, then
amplified, and finally the mean squared fluctuation (the noise power)
is measured.
\label{FQHE}
}
\end{figure}

A first example of $q\neq e$ is the shot noise at a tunnel junction
between a normal metal and a superconductor. Charge is added to the
superconductor in Cooper pairs, so one expects $q=2e$ and $F=2$. This
doubling of the Poisson noise has been measured very recently.\cite{Lef02}
(Earlier experiments \cite{Koz00} in a disordered system will be discussed
later on.)

A second example is offered by the fractional quantum Hall effect. It
is a non-trivial implication of Robert Laughlin's theory that tunneling
from one edge of a Hall bar to the opposite edge proceeds in units of a
fraction $q=e/(2p+1)$ of the elementary charge.\cite{Kan94} The integer
$p$ is determined by the filling fraction $p/(2p+1)$ of the lowest Landau
level. Christian Glattli and collaborators of the Centre d'\'{E}tudes de
Saclay in France and Michael Reznikov and collaborators of the Weizmann
Institute in Israel independently measured $F=1/3$ in the fractional
quantum Hall effect\cite{Sam97} (see figure \ref{FQHE}). More
recently, the Weizmann group extended the noise measurements to $p=2$
and $p=3$. The experiments at $p=2$ show that the charge inferred from
the noise may be a multiple of $e/(2p+1)$ at the lowest temperatures,
as if the quasiparticles tunnel in bunches. How to explain
this bunching is still unknown.

\section*{\large\bf Quiet electrons}
\noindent
Correlations reduce the noise below the value 
\begin{equation} 
S_{\rm Poisson}=2e\bar{I}\label{SPoisson} 
\end{equation} 
expected for a Poisson process of uncorrelated current pulses of
charge $q=e$. Coulomb repulsion is one source of correlations, but it
is strongly screened in a metal and ineffective.  The dominant source of
correlations is the Pauli principle, which prevents double occupancy of an
electronic state and leads to Fermi statistics in thermal equilibrium. In
a vacuum tube or tunnel junction the mean occupation of a state is so
small that the Pauli principle is inoperative (and Fermi statistics
is indistinguishable from Boltzmann statistics), but this is not so in
a metal.

An efficient way of accounting for the correlations uses Landauer's
description of electrical conduction as a transmission problem. According
to the Landauer formula, the time-averaged current $\bar{I}$ equals
the conductance quantum $2e^{2}/h$ (including a factor of two for
spin), times the applied voltage $V$, times the sum over transmission
probabilities $T_{n}$:
\begin{equation}
\bar{I}=\frac{2e^{2}}{h}V\sum_{n=1}^{N}T_{n}.\label{Gformula}
\end{equation}
The conductor can be viewed as a parallel circuit of $N$ independent
transmission channels with a channel-dependent transmission probability
$T_{n}$.  Formally, the $T_{n}$'s are defined as the eigenvalues of the
product $t\cdot t^{\dagger}$ of the $N\times N$ transmission matrix $t$
and its Hermitian conjugate.  In a one-dimensional conductor, which
by definition has one channel, one would have simply $T_{1}=|t|^{2}$,
with $t$ the transmission amplitude.

The number of channels $N$ is a large number in a typical metal wire.
One has $N\simeq A/\lambda_{F}^{2}$ up to a numerical coefficient for a
wire with cross-sectional area $A$ and Fermi wave length $\lambda_{F}$.
Due to the small Fermi wave length $\lambda_{F}\simeq 1\,{\rm \AA}$
of a metal, $N$ is of order $10^{7}$ for a typical metal wire of width
$1\,\mu{\rm m}$ and thickness $100\,{\rm nm}$. In a semiconductor typical
values of $N$ are smaller but still $\gg 1$.

At zero temperature the noise is related to the transmission probabilities by\cite{Khl87}
\begin{equation}
S=2e\frac{2e^{2}}{h}V\sum_{n=1}^{N}T_{n}(1-T_{n}).\label{Pformula}
\end{equation}
The factor $1-T_{n}$ describes the reduction of noise due to the Pauli
principle. Without it, one would have simply $S=S_{\rm Poisson}$.

The shot noise formula (\ref{Pformula}) has an instructive statistical
interpretation.\cite{Lev93}  Consider first a one-dimensional
conductor. Electrons in a range $eV$ above the Fermi level enter the
conductor at a rate $eV/h$. In a time $\tau$ the number of attempted
transmissions is $\tau eV/h$. There are no fluctuations in this number
at zero temperature, since each occupied state contains exactly one
electron (Pauli principle). Fluctuations in the transmitted charge $Q$
arise because the transmission attempts are succesful with a probability
$T_{1}$ which is different from 0 or 1. The statistics of $Q$ is {\em
binomial}, just as the statistics of the number of heads when tossing a
coin. The mean-squared fluctuation $\langle\delta Q^{2}\rangle$ of the
charge for binomial statistics is given by
\begin{equation}
\langle\delta Q^{2}\rangle =e^{2}(\tau eV/h)T_{1}(1-T_{1}).\label{varnbinomial}
\end{equation}
The relation $S=(2/\tau)\langle\delta Q^{2}\rangle$ between the
mean-squared fluctuation of the current and of the transmitted
charge brings us to eq.\ (\ref{Pformula}) for a single channel. Since
fluctuations in different channels are independent, the multi-channel
version is simply a sum over channels.

The quantum shot noise formula (\ref{Pformula}) has been tested
experimentally in a variety of systems. The groups of Reznikov and Glattli
used a quantum point contact: A narrow constriction in a two-dimensional
electron gas with a quantized conductance. The quantization occurs because
the transmission probabilities are either close to 0 or close to 1. Eq.\
(\ref{Pformula}) predicts that the shot noise should vanish when the
conductance is quantized, and this was indeed observed. (The experiment
was reviewed by Henk van Houten and Beenakker in {\sc Physics Today},
July 1996, page 22.)

A more stringent test used a single-atom junction, obtained by the
controlled breaking of a thin aluminum wire.\cite{Cro01} The junction
is so narrow that the entire current is carried by only three channels
($N=3$).  The transmission probabilities $T_{1},T_{2},T_{3}$ could be
measured independently from the current--voltage characteristic in the
superconducting state of aluminum. By inserting these three numbers (the
``pin code'' of the junction) into eq.\ (\ref{Pformula}), a theoretical
prediction is obtained for the shot noise power --- which turned out to
be in good agreement with the measured value.

\section*{\large\bf Detecting open transmission channels}
\noindent

The analogy between an electron emitted by a cathode and a bullet shot
by a gun works well for a vacuum tube or a point contact, but seems
a rather naive description of the electrical current in a disordered
metal or semiconductor. There is no identifiable emission event when
current flows through a metal and one might question the very existence
of shot noise. Indeed, for three quarters of a century after the first
vacuum tube experiments there did not exist a single measurement of shot
noise in a metal. A macroscopic conductor (say, a piece of copper wire)
shows thermal noise, but no shot noise.

We now understand that the basic requirement on length scale and
temperature is that the length $L$ of the wire should be short compared
to the inelastic electron-phonon scattering length $l_{\rm in}$, which
becomes longer and longer as one lowers the temperature.  For $L>l_{\rm
in}$ each segment of the wire of length $l_{\rm in}$ generates independent
voltage fluctuations, and the net result is that the shot noise power is
reduced by a factor $l_{\rm in}/L$. Thermal fluctuations, in contrast, are
not reduced by inelastic scattering (which can only help the establishment
of thermal equilibrium).  This explains why only thermal noise could
be observed in macroscopic conductors. (As an aside, we mention that
inelastic electron-electron scattering, which persists until much lower
temperatures than electron-phonon scattering, does not suppress shot
noise, but rather enhances the noise power a little bit.\cite{Nag95})

\begin{figure}
\includegraphics[width=.8\linewidth]{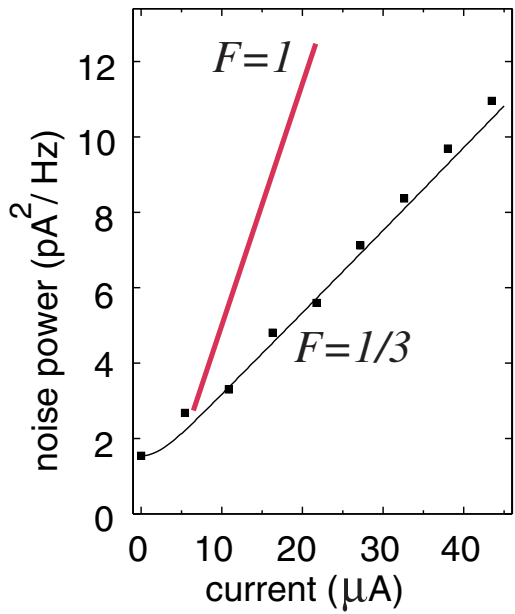}
\caption{
Sub-Poissonian shot noise in a disordered gold wire (dimensions 940~nm
$\times$ 100~nm). At low currents the noise saturates at the level set
by the temperature of 0.3~K. [Adapted from M. Henny et al.\protect\cite{Ste96}] \label{henny}
}
\end{figure}

Early experiments\cite{Lie94} on mesoscopic semiconducting wires observed the
linear relation between noise power and current that is the signature
of shot noise, but could not accurately measure the slope. The first
quantitative measurement was performed in a thin-film silver wire by
Andrew Steinbach and John Martinis at the US National Institute of
Standards and Technology in Boulder, collaborating with Michel Devoret
from Saclay.\cite{Ste96} 

The data shown in figure \ref{henny} (from a more recent experiment)
presents a puzzle: If we calculate the slope, we find a Fano factor of
1/3 rather than 1. Surely there are no fractional charges in a normal
metal conductor?

A one-third Fano factor in a disordered conductor had actually been
predicted prior to the experiments. The prediction was made independently
by Kirill Nagaev of the Institute of Radio-Engineering and Electronics
in Moscow and by one of the authors (Beenakker) with Markus B\"{u}ttiker
of the University of Geneva.\cite{Bee92} To understand the experimental
finding we recall the general shot noise formula (\ref{Pformula}), which
tells us that sub-Poissonian noise ($F<1$) occurs when some channels
are {\em not\/} weakly transmitted. These socalled ``open channels''
have $T_{n}$ close to 1 and therefore contribute less to the noise than
expected for a Poisson process.

\begin{figure}
\includegraphics[width=.8\linewidth]{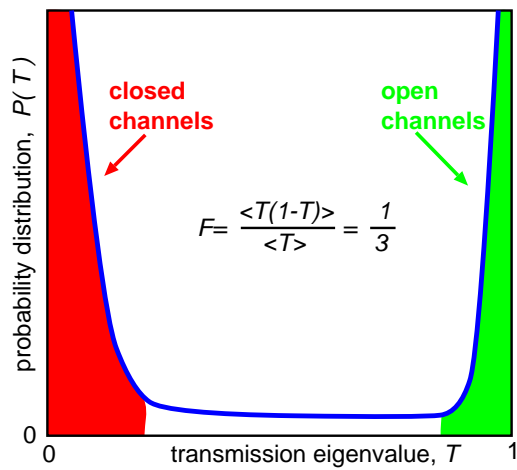}
\caption{
Bimodal probability distribution of the transmission eigenvalues,
with a peak at 0 (closed channels) and a peak at 1 (open channels). The
functional form of the distribution (derived by Dorokhov) is $P(T)\propto
T^{-1}(1-T)^{-1/2}$, with a mean-free-path dependent cutoff at
exponentially small $T$. The one-third Fano factor follows directly from
the ratio $\int T^{2}P(T)dT/\int TP(T)dT=2/3$. The cutoff affects only
the normalization of $P(T)$ and drops out of this ratio, which takes on
a universal value.
\label{bimodal}
}
\end{figure}

The appearance of open channels in a disordered conductor is surprising.
Oleg Dorokhov of the Landau Institute in Moscow first noticed the
existence of open channels in 1984, but the physical implications were
only understood some years later, notably through the work of Yoseph
Imry of the Weizmann Institute. The one-third Fano factor
follows directly from the probability distribution of the transmission
eigenvalues, see figure \ref{bimodal}.

We conclude this section by referring to the experimental
demonstrations\cite{Koz00} of the interplay between the doubling of shot
noise due to superconductivity and the 1/3 reduction due to open channels,
resulting in a 2/3 Fano factor. These experiments show that open
channels are a general and universal property of disordered systems.

\section*{\large\bf Distinguishing particles from waves}
\noindent
So far we have encountered two diagnostic properties of shot noise:
It measures the unit of transferred charge in a tunnel junction and
it detects open transmission channels in a disordered wire. A third
diagnostic appears in semiconductor microcavities known as quantum
dots or electron billiards. These are small confined regions in a
two-dimensional electron gas, free of disorder, with two narrow openings
through which a current is passed. If the shape of the confining potential
is sufficiently irregular (which it typically is), the classical dynamics
is chaotic and one can search for traces of this chaos in the quantum
mechanical properties.  This is the field of quantum chaos.

Here is the third diagnostic: Shot noise in an electron billiard can
distinguish deterministic scattering, characteristic for particles, from
stochastic scattering, characteristic for waves.  Particle dynamics
is deterministic: A given initial position and momentum fixes the
entire trajectory. In particular, it fixes whether the particle will
be transmitted or reflected, so the scattering is noiseless on all time
scales. Wave dynamics is stochastic: The quantum uncertainty in position
and momentum introduces a probabilistic element into the dynamics,
so it becomes noisy on sufficiently long time scales.

The suppression of shot noise in a conductor with deterministic scattering
was predicted many years ago\cite{Bee91} from this qualitative argument. A
better understanding, and a quantitative description, of how shot noise
measures the transition from particle to wave dynamics was developed
recently by Oded Agam of the Hebrew University in Jerusalem, Igor Aleiner
of the State University of New York in Stony Brook, and Anatoly Larkin
of the University of Minnesota in Minneapolis.\cite{Aga00} The key
concept is the Ehrenfest time, which is the characteristic time scale
of quantum chaos.

In classical chaos, the trajectories are highly sensitive to small
changes in the initial conditions (although uniquely determined by
them). A change $\delta x(0)$ in the initial coordinate is amplified
exponentially in time: $\delta x(t)=\delta x(0)e^{\alpha t}$.  Quantum
mechanics introduces an uncertainty in $\delta x(0)$ of the order of
the Fermi wave length $\lambda_{F}$.  One can think of $\delta x(0)$
as the initial size of a wave packet. The wave packet spreads over the
entire billiard (of size $L$) when $\delta x(t)=L$. The time
\begin{equation}
\tau_{E}=\alpha^{-1}\ln(L/\lambda_{F})\label{tauEdef}
\end{equation}
at which this happens is called the Ehrenfest time.

\begin{figure}
\includegraphics[width=.8\linewidth]{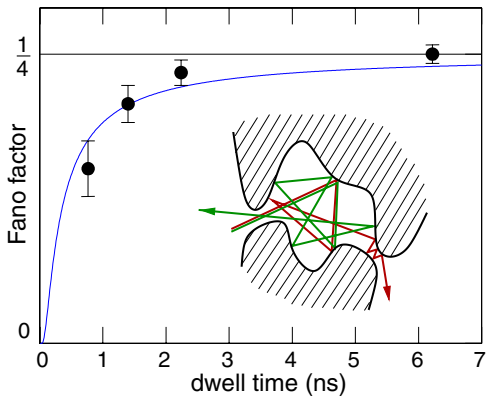}
\caption{
Dependence of the Fano factor $F$ of an electron billiard on the average
time $\tau_{\rm dwell}$ that an electron dwells inside. The data points
with error bars are measured in a two-dimensional electron gas, the solid
curve is the theoretical prediction $F=\frac{1}{4}\exp(-\tau_{E}/\tau_{\rm
dwell})$ for the transition from stochastic to deterministic scattering
(with Ehrenfest time $\tau_{E}=0.27\,{\rm ns}$ as a fit parameter).
The inset shows graphically the sensitivity to initial conditions of
the chaotic dynamics.  (Adapted from ref.\ \onlinecite{Aga00}, with
experimental data from ref.\ \onlinecite{Obe02}.)
\label{agam}
}
\end{figure}

The name refers to Paul Ehrenfest's 1927 principle that quantum mechanical
wave packets follow classical, deterministic, equations of motion. In
quantum chaos this correspondence principle loses its meaning (and the
dynamics becomes stochastic) on time scales greater than $\tau_{E}$. An
electron entering the billiard through one of the openings dwells inside
on average for a time $\tau_{\rm dwell}$ before exiting again. Whether
the dynamics is deterministic or stochastic depends, therefore, on the
ratio $\tau_{\rm dwell}/\tau_{E}$.  The theoretical expectation for
the dependence of the Fano factor on this ratio is plotted in figure
\ref{agam}.

An experimental search for the suppression of shot noise by
deterministic scattering was carried out at the University of Basel
by Stefan Oberholzer, Eugene Sukhorukov, and one of the authors
(Sch\"{o}nenberger).\cite{Obe02} The data is included in figure
\ref{agam}. An electron billiard (area $A\approx 53\,\mu{\rm m}^{2}$) with
two openings of variable width was created in a two-dimensional electron
gas by means of gate electrodes. The dwell time (given by $\tau_{\rm
dwell}=m^{\ast}A/\hbar N$, with $m^{\ast}$ the electron effective mass)
was varied by changing the number of modes $N$ transmitted through each
of the openings.

The Fano factor has the value $1/4$ for long dwell times, as expected
for stochastic chaotic scattering. The $1/4$ Fano factor for a chaotic
billiard has the same origin as the $1/3$ Fano factor for a disordered
wire, explained in figure \ref{bimodal}. (The different number results
because of a larger fraction of open channels in a billiard geometry.)
The reduction of the Fano factor below $1/4$ at shorter dwell times
fits the exponential function $F=\frac{1}{4}\exp(-\tau_{E}/\tau_{\rm
dwell})$ of Agam, Aleiner, and Larkin. However, the accuracy and
range of the experimental data is not yet sufficient to distinguish
this prediction from competing theories (notably the rational function
$F=\frac{1}{4}(1+\tau_{E}/\tau_{\rm dwell})^{-1}$ predicted by Sukhorukov
for short-range impurity scattering).

\section*{\large\bf Entanglement detector}
\noindent
The fourth and final diagnostic property that we would like to discuss,
shot noise as detector of entanglement, was proposed by Sukhorukov with
Guido Burkard and Daniel Loss from the University of Basel.\cite{Bur00}

A multi-particle state is entangled if it can not be factorized into a
product of single-particle states. Entanglement is the primary resource
in quantum computing, in the sense that any speed-up relative to a
classical computer vanishes if the entanglement is lost, typically through
interaction with the environment (see the article by John Preskill, {\sc
Physics Today}, July 1999, page 24). Electron-electron interactions lead
quite naturally to an entangled state, but in order to make use of the
entanglement in a computation one would need to be able to spatially
separate the electrons without destroying the entanglement. In this
respect the situation in the solid state is opposite to that in quantum
optics, where the production of entangled photons is a complex operation,
while their spatial separation is easy.

\begin{figure}
\includegraphics[width=.8\linewidth]{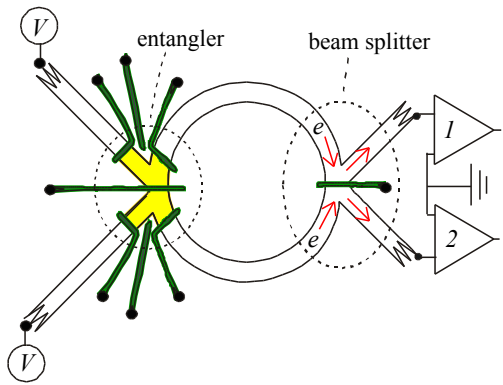}
\caption{
Proposal for production and detection of a spin-entangled electron
pair. The double quantum dot (shown in yellow) is defined by gate
electrodes (green) on a two-dimensional electron gas. The two voltage
sources at the far left inject one electron in each dot, resulting in an
entangled spin-singlet ground state. A voltage pulse on the gates then
forces the two electrons to enter opposite arms of the ring. Scattering
of the electron pair by a tunnel barrier (red arrows) creates shot
noise in each of the two outgoing leads at the far right, which is
measured by a pair of amplifiers (1,2). The observation of a positive
correlation between the current fluctuations at 1 and 2 is a signature of
the entangled spin-singlet state. (Figure courtesy of L.P. Kouwenhoven and A.F. Morpurgo, Delft University of Technology.)
\label{leo_entangler}
}
\end{figure}

One road towards a solid-state based quantum computer has as its
building block a pair of quantum dots, each containing a single
electron. The strong Coulomb repulsion keeps the electrons separate,
as desired. In the ground state the two spins are entangled
in the singlet state $|\!\uparrow\rangle|\!\downarrow\rangle
-|\!\downarrow\rangle|\!\uparrow\rangle$. This state may already have
been realized experimentally,\cite{Hol02} but how can one tell? Noise
has the answer.

To appreciate this we contrast ``quiet electrons'' with ``noisy
photons''. We recall that Fermi statistics causes the electron noise
to be smaller than the Poisson value (\ref{SPoisson}) expected for
classical particles. For photons the noise is bigger than the Poisson
value because of Bose statistics. What distinghuishes the two is whether
the wave function is symmetric or antisymmetric under exchange of particle
coordinates.  A symmetric wave function causes the particles to bunch
together, increasing the noise, while an antisymmetric wave function has
the opposite effect (``antibunching''). The key point here is that only
the symmetry of the {\em spatial\/} part of the wave function matters for
the noise. Although the full many-body electron wavefunction, including
the spin degrees of freedom, is always antisymmetric, the spatial part is
not so constrained. In particular, electrons in the spin-singlet state
have a symmetric wave function with respect to exchange of coordinates,
and will therefore bunch together like photons.

The experiment proposed by the Basel theorists is sketched in
figure \ref{leo_entangler}. The two building blocks are the {\em
entangler\/} and the {\em beam splitter\/}. The beam splitter is used
to perform the electronic analogue of the optical Hanbury Brown and
Twiss experiment.\cite{Hen99} In such an experiment one measures the
cross-correlation of the current fluctuations in the two arms of a beam
splitter. Without entanglement, the correlation is positive for photons
(bunching) and negative for electrons (antibunching). The observation
of a positive correlation for electrons is a signature of the entangled
spin-singlet state. In a statistical sense, the entanglement makes the
electrons behave as photons.

An alternative to the proposal shown in figure \ref{leo_entangler}
is to start from Cooper pairs in a superconductor, which are also in a
spin-singlet state.\cite{Bur00} The Cooper pairs can be extracted from
the superconductor and injected into a normal metal by application of
a voltage over a tunnel barrier at the metal--superconductor interface.

Experimental realization of one these theoretical proposals would open
up a new chapter in the use of noise as a probe of quantum mechanical
properties of electrons. Although this range of applications is still
in its infancy, the field as a whole has progressed far enough to prove
Landauer right: There {\em is\/} a signal in the noise.

\end{document}